\documentclass[reprint,superscriptaddress, amsmath, amssymb, prd, aps, showkeys]{revtex4-1}

\usepackage{subfig, caption}
\usepackage{natbib}
\usepackage{mathtools}
\usepackage{txfonts}
\usepackage{scalerel}
\usepackage{graphicx}  
\usepackage{bm}        
\usepackage{amssymb}   
\usepackage{lipsum}
\usepackage{lineno}
\usepackage{amsmath,array}
\usepackage{caption}
\usepackage{hyperref}
\usepackage{enumitem}
\hypersetup{
    colorlinks=true,
    linkcolor=blue,
    filecolor=magneta,      
    urlcolor=blue,
}
\usepackage{scalerel}
\usepackage{tikz}
\usetikzlibrary{svg.path}
\definecolor{orcidlogocol}{HTML}{A6CE39}
\tikzset{
    orcidlogo/.pic={
        \fill[orcidlogocol] svg{M256,128c0,70.7-57.3,128-128,128C57.3,256,0,198.7,0,128C0,57.3,57.3,0,128,0C198.7,0,256,57.3,256,128z};
        \fill[white] svg{M86.3,186.2H70.9V79.1h15.4v48.4V186.2z}
        svg{M108.9,79.1h41.6c39.6,0,57,28.3,57,53.6c0,27.5-21.5,53.6-56.8,53.6h-41.8V79.1z M124.3,172.4h24.5c34.9,0,42.9-26.5,42.9-39.7c0-21.5-13.7-39.7-43.7-39.7h-23.7V172.4z}
        svg{M88.7,56.8c0,5.5-4.5,10.1-10.1,10.1c-5.6,0-10.1-4.6-10.1-10.1c0-5.6,4.5-10.1,10.1-10.1C84.2,46.7,88.7,51.3,88.7,56.8z};
    }
}
\newcommand\orcidicon[1]{\href{https://orcid.org/#1}{\mbox{\scalerel*{
                \begin{tikzpicture}[yscale=-1,transform shape]
                \pic{orcidlogo};
                \end{tikzpicture}
            }{|}}}}
\DeclareCaptionJustification{justified}{\leftskip=0pt \rightskip=0pt \parfillskip=0pt plus 1fil}

\begin{document}
\title{Rotating Scalar Field and Formation of Bose Stars}
\author{Kuldeep J. Purohit}
\email{kuldeepjrajpurohit@gmail.com}
\affiliation{%
    Physics Department, Faculty of Science,
    The Maharaja Sayajirao University of Baroda, Vadodara 390 002, Gujarat, India}
\author{Pravin Kumar Natwariya $^{\orcidicon{0000-0001-9072-8430}}$\,}
\email{pvn.sps@gmail.com, pravin@prl.res.in}
\affiliation{%
    Physical Research Laboratory, Theoretical Physics Division, Ahmedabad, Gujarat 380 009, India}
\affiliation{%
    Department of Physics, Indian Institute of Technology, Gandhinagar, Palaj, Gujarat 382 355, India}  
\author{Jitesh R. Bhatt $^{\orcidicon{0000-0001-7465-8292}}$\,}
\email{jeet@prl.res.in}
\affiliation{%
    Physical Research Laboratory, Theoretical Physics Division, Ahmedabad, Gujarat 380 009, India}
\author{Prashant K. Mehta}
\email{pk.mehta-phy@msubaroda.co.in}
\affiliation{%
   Physics Department, Faculty of Science,
   The Maharaja Sayajirao University of Baroda, Vadodara 390 002, Gujarat, India}

\date{\today}

\begin{abstract}
{\centering \bf Abstract\par}
We study numerical evolutions of an initial cloud of self-gravitating bosonic dark matter with finite angular momentum and self-interaction in kinetic regime. It is demonstrated that such a system can undergo gravitational condensation and form a Bose star. The results show that the gravitational condensation time is strongly influenced by the presence of finite angular momentum or the strength of self-interaction.  We find that in the cases related with attractive or no self-interaction, there is no significant transfer of angular momentum from the initial cloud to the formed star. However, for the case repulsive interaction our results indicate that such a angular-momentum transfer is possible. These results are consistent with the earlier analytical work where the stability of the rotating boson star was considered \cite{Dmitriev_2021}.

\end{abstract}

\maketitle

\section{Introduction}

The observation of gravitational-wave signal GW190521 by advanced LIGO/Virgo collaboration \cite{Abbott_2020} has opened up an interesting possibility of detecting exotic compact objects like Bose stars in the Universe \cite{Bustillo_2021, Dmitriev_2021, Levkov_2018,Siemonsen_2021, chen:2021}. GW190521 can be a potential signal from the merger of two Bose stars \cite{Bustillo_2021}.
It ought to be noted here that Bose stars have been the subject of investigations much longer before the advanced LIGO/Virgo collaboration (\cite{Wheeler:1955}, for a review see \cite{Liebling_2017}).
Among the different models of these exotic compact objects, the scalar field Bose star is one of the most studied ones \cite{Liebling_2017}. Since dark-matter constitute around 83\% of the matter in the Universe, the boson star may have been made of dark matter. A Scalar degree of freedom arises naturally in the fundamental theories related to physics beyond the Standard model of particle physics. Many of these new scalars can be good candidates for dark matter
like QCD-axion \cite{Preskill:1983, Kolb_1993, Turner:1986, Hogan:1988} , fuzzy cold dark matter \cite{Hu:2000} etc.
Thus it is possible that an initial self-gravitating cloud of the dark scalar boson can form a macroscopic bound state
in the Universe. If the self-gravitating bosons are sufficiently cold then they can form a Bose-Einstein condensate and the bound state is called the Bose Star \cite{Levkov_2018, Colpi:1986}. If the thermal de-Borglie wavelength $\lambda_{dB}=[2\,\pi/(m\,T)]^{1/2}$, is much larger than the inter-particle spacing
$n^{-1/3} $, where, $m\ , T \ \& \ n$ respectively denote mass, temperature and number density of the particles, then the Bose-Einstein condensation may occur. This inequality can lead to the following upper bound on the mass of the scalar dark-matter candidate, i.e. $m\,<\,1.87$~eV \cite{Fukuyama_2008, Bohmer_2007, Chavanis_2011}.

There exist various analytic solutions to the Einstein-Klein-Gordon equations describing properties of a Bose star \cite{marsh:2021, RUFFINI:1969}. Although the study of Bose stars started long ago, there is a little progress in understanding the evolution and formation of the Bose star under various physical situations, for e.g. attractive/repulsive self-interaction between dark matter particles \cite{Eby_2016, Madsen:1990}. 

However, the recent works \cite{Levkov_2018, Veltmaat_2018} has highlighted that the importance of studying how the various state of initial field can influence the star forming processes.

 In reference \cite{Levkov_2018} the authors demonstrated that how a randomly distributed low density(kinetic regime) initial clould of non-interacting bosons can undergo a gravitational collapse to form a Bose star.
  In kinetic regime, the particle de Broglie wavelength $(mv)$ and time scale $(mv^2)$ satisfy the following conditions:
\begin{equation}
	mvR \gg 1, \ \ \ \ \ \ \ \  mv^2 \tau_{gr}\gg1
\end{equation}
where, $m$ is the mass of the Dark Matter particle, $v$ is its characteristic velocity, R is the halo size, and $\tau_{gr}$ is the condensation time. $\tau_{gr}$ is the time where Bose star formation begins.
The authors argue that their study  can help in solving the missing satellite problem and shed light on Fast Radio Bursts \cite{Tkachev:2014}, ARCADE 2 \cite{Fixsen2011} and EDGES \cite{Bowman:2018yin} observations. The role of self-interaction in the formation of Bose stars has been studied in reference \cite{Veltmaat_2018}. Authors of the reference \cite{chen:2021}, study the formation and stability of Bose stars by considering self-interaction---both attractive and repulsive.

\begin{figure*}
	\centering
	\begin{tabular}{@{}c@{}}
		\includegraphics[width=18cm,height=5cm]{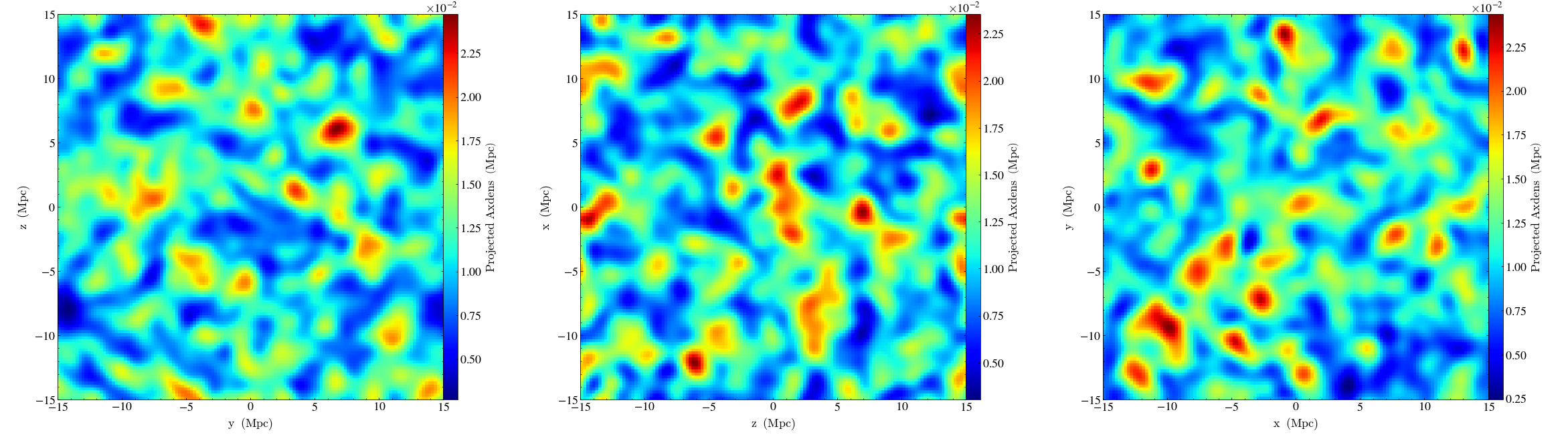} \\[\abovecaptionskip]
	\end{tabular}	
	
	
	\begin{tabular}{@{}c@{}}
		\includegraphics[width=18cm,height=5cm]{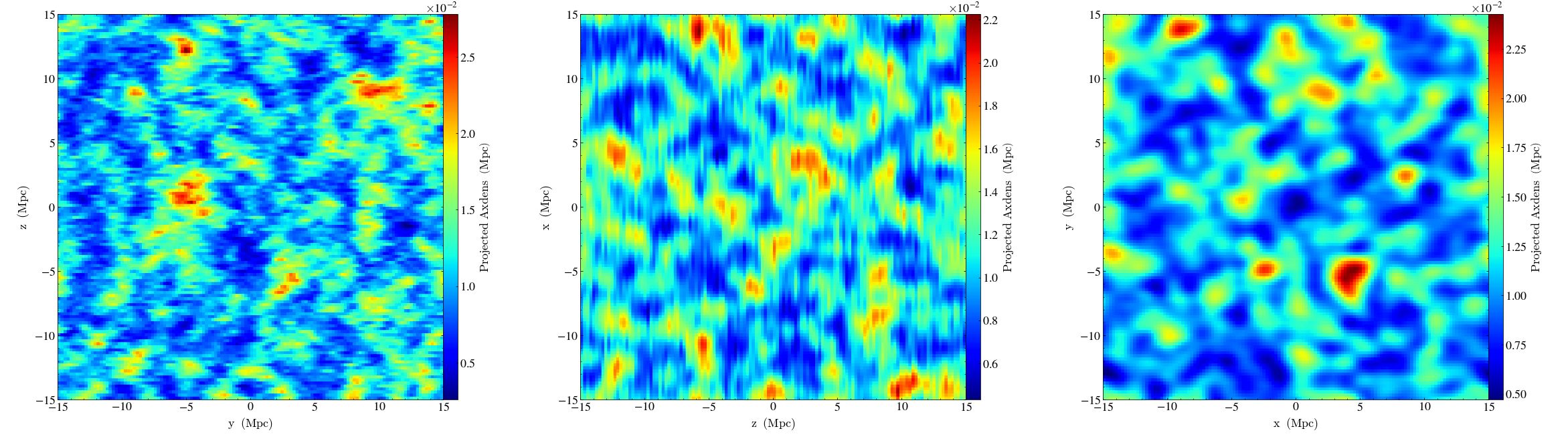} \\[\abovecaptionskip]
	\end{tabular}

	\caption{The initial snapshots of density profiles of $|\tilde{\psi}|^2$ at the initial time ($\tilde t=0$) in $\tilde y \tilde z$, $\tilde z \tilde x$ and $\tilde x\tilde y$ planes. The top panel from left to right describes the initial cloud density for $\mathcal{\tilde L_{\rm tot}}=0$. The middle and bottom panels respectively correspond to $\mathcal{\tilde L_{\rm tot}}=3$ and 5 cases.}\label{plot1}
\end{figure*}

Gravitational shearing plays a crucial role in the generation of rotation in gravitational clouds. In recent work, for example, velocity gradients were observed across the molecular clouds of the M33 galaxy to study their rotation, and they were found rotating \cite{Larson:1948, Seigar:2005}. Therefore it is very essential to study the formation of Bose stars under such a scenario. In the present study, we intend to study the effect of angular momentum present in the initial bosonic cloud undergoing a gravitational collapse. It becomes interesting to study whether such rotating clouds eventually form a rotating Bose star or not. In the literature, there exist several studies where the authors analyzed Bose stars with finite angular momentum. Recently, the authors of the reference \cite{Dmitriev_2021}, have shown analytically that in the absence of self-interaction or finite attractive interaction Bose stars with finite angular momentum are unstable. However, the rotating Bose star configurations can be possible when the repulsive self-interaction is present. In reference \cite{Siemonsen_2021},  the authors study the stability of rotating Bose stars in various massive scalar field models described by the Klein-Gordon equation. Here it is found that mini Bose star configurations are unstable and the non-axisymmetric instability can grow exponentially. However, the instability vanishes when the growth rate $\omega_I$ is comparable to the mass of the scalar field. In the reference  \cite{Gual_2019}, authors consider the nonlinear Einstein Klein-Gordon and Einstein Proca systems for initially spinning axisymmetric clouds, and investigate the formation and stability of rotating Bose stars under the self-gravitating field. For the scalar case, nonaxisymmetric instability always occurs and ejects all the angular momentum from the star. But for the Proca system, one can find a stable configuration of the rotating star. 

\begin{figure*}
	\centering
		\includegraphics[width=18cm,height=10cm]{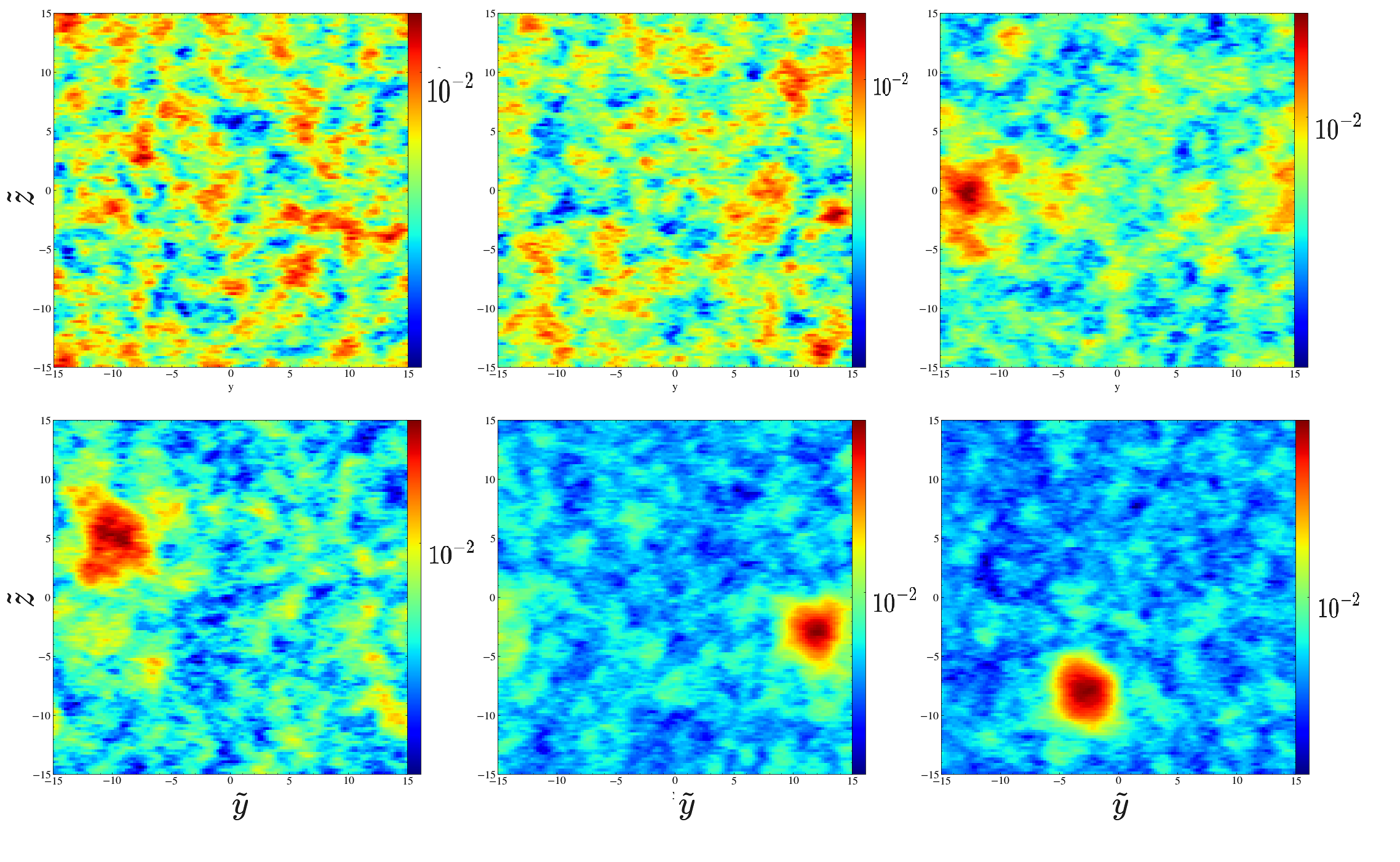}
	\caption{Snapshots of $|\tilde{\psi}|^2$ at different values of time are shown in $\tilde{x}\tilde{y}$-plane. All the plots represent the case when the initial cloud has total-angular momentum $\mathcal{\tilde L_{\rm tot}}=5$ 
		and no self-interaction $\tilde{g}=0$.  The gravitational condensation time $\tilde{\tau}_{gr}$ for this case is around 15600. The plots in the upper panel from the left to right respectively correspond to time $\tilde{t}=0,\, 0.5\tau_{gr},\, \& 1\tau_{gr}$. Three plots in the bottom correspond to $1.15\tilde{\tau}_{gr},\,1.15\tilde{\tau}_{gr}\,\& 1.8\tilde{\tau}_{gr}$.}
\end{figure*}

In the present work, we intend to study the formation of a Bose star when the initial field has a finite angular momentum. The initial cloud is considered to be dilute nonaxisymmetric but with a nonzero angular momentum. We study the evolution of such a system under the influence of its self-gravity. We also analyzed the effect of both attractive and repulsive interactions on star formation. The presence of initial field rotation can significantly affect the star formation time for different values of angular momentum. We shall demonstrate below that the presence of angular momentum and self-interaction can strongly influence the dynamics of Bose star formation and the properties of the star.

This work is divided into the following sections: In section [\ref{sec2}], we discuss the time evolution of boson particles for various cases (attractive/repulsive self-interaction and angular momentum); in section [\ref{sec3}], we discuss the results for above-mentioned cases, and finally, we conclude the results in section [\ref{sec4}]. Throughout the paper, we use natural units $c=k_B=\hbar=1$. Where $c$ is the speed of light, $k_B$ is the Boltzmann constant and $\hbar$ is the reduced Planck constant. The quantities in the boldface represent the three vectors.

\section{Evolution of the Bose star}\label{sec2}
In the Non-relativistic limit and for high mean occupation number, the system can be represented by a classical scalar field, $\psi(\bm r,t)$, whose time evolution can be described by the Gross-Pitaevskii-Poisson  equations
\cite{chen:2021, Chavanis_:2011, Bohmer_:2007, Eby_2016},
\begin{alignat}{2}
i\,\frac{\partial \psi}{\partial t}&=-\frac{1}{2\,m}\,\nabla^2\psi+m\,\Phi\,\psi+{g}\,|\psi|^2\,\psi \,,\label{eq1}\\
\nabla^2\Phi &= 4\,\pi\, G\, m\,(\,|\psi|^2-n\, )\,,\label{eq2}
\end{alignat}
where,  $\Phi\equiv\Phi(\bm r,t)$ is the gravitational potential. $G$ is the universal gravitational constant and $n$ is the mean particle density of boson gas.  \(g\) is the self-interaction coupling constant, it is positive for repulsive self-interaction and negative for attractive self-interaction. $g=0$ corresponds to a non-interacting case. For a box with size, $L$ and a total number of particles, $N$, the number density is $n=N/L^3$. The above equations \eqref{eq1} and \eqref{eq2}, can be written in dimensionless form using: $r=(1/m v_0)\,\tilde{r}\,$, $t=(1/m\,v_0^2)\,\tilde{t}\,$, $\Phi=v_0^2\,\tilde{\Phi}\,$, $\psi=\big(v_0^2\sqrt{m/4\pi G}\,\big)\,\tilde{\psi}$ and $g=(4\pi\, G/v_0^2)\,\tilde{g}\,$.
\begin{alignat}{2}
i\,\frac{\partial}{\partial{\tilde{t}}}\,\tilde{\psi}&=-\frac{1}{2}\,\tilde{\nabla}^2\,\tilde{\psi} + \tilde{\Phi}\,\tilde{\psi}+\tilde{g}\,|\tilde{\psi}|^2\,\tilde{\psi},
\label{eq3}\\
\tilde{\nabla}^2\,{\tilde{\Phi}}&=|\tilde{\psi}|^2-\tilde{n},\ 
\label{eq4}
\end{alignat}    

where $\tilde{n} = (4\pi G/mv_0^4)$ , $N=\int d^3r\, |\psi|^2$, one can get $N=(v_0/4\pi Gm^2)\,\tilde N$. Here, $v_0$ is the reference velocity \cite{chen:2021}

Next, we specify the initial conditions in the Fourier space. First, consider the case when there is no angular momentum, we choose  a "Gaussian" wave function,
\begin{equation}
\tilde{\psi}_p\,=\,\frac{\tilde{ N}^{1/2}}{\pi^{3/4}}e^{-\tilde{p}^2/2} e^{i\alpha_p},\label{init}
\end{equation}
\noindent where $\alpha_p$  is a random phase distributed over the Fourier space between the number 0 to $2\pi$. This initial form of the distribution function used as an input to the numerical code \texttt{AxioNyx} \footnote{\href{https://github.com/axionyx}{https://github.com/axionyx}} with necessary modifications \cite{Schwabe_2020, 
	Almgren_2013}.  Next, if the above function is  Fourier transformed in the coordinate space, then the random phase inclusion would make a profile of the amplitude $\tilde{\psi}(\tilde{x},\tilde{y},\tilde{z}, \tilde{t}=0)$ homogeneous and isotropic \cite{Levkov_2018}. 

 For all the numerical results presented in this work, the following prescription to generate angular momentum in the initial wave function: We consider the function,
\begin{alignat}{2}
\tilde\psi({\tilde{\bm r}},\tilde{t}=0)\,=\,\frac{\tilde{ N}^{1/2}}{\pi^{3/4}}\, e^{-\tilde{\bm r}^2/2}\ e^{i\,l\,\tan^{-1}(\tilde y/\tilde x)}\,,\label{eq5}
\end{alignat}
\noindent
The initial wave function which enters the GPP equations( \eqref{eq3}, \eqref{eq4}) is obtained first by transforming the function in equation(7) and multiplying it with $\alpha_p$ (the random-phase) and back transforming the product to the configuration space. Here we note that we have tried various other prescriptions for introducing non-zero angular momentum in the initial cloud also. However, the change in the angular momentum does not change the numerical results in any significant way.

  Further, we would like to note here that we have chosen the periodic boundary conditions as is the case in the earlier works  \cite{Schwabe_2020, Levkov_2018, chen:2021}. This means that if the box length is $L$, then the wave function $\psi(x)$ needs to satisfy the following condition: $\psi(x+L)=\psi(x)$. Since the interaction term in the Hamiltonian density $\tilde{g}\,|\tilde{\psi}({\bm \tilde{x}}, \tilde{t})|^2$  depends only on the local coordinate ${\bm x}$, one may not be required to introduce a cut-off to remove the effect from the particles outside the box. Thus one expects that the numerical results presented here may not have any effect due finite size of the box.

The stress-energy tensor can be written as \cite{Weinberg:1972}
\begin{alignat}{2}
\widetilde  T^{0i}=\frac{i}{2}\left( \tilde \psi \tilde \partial_i \tilde \psi^*-\tilde \psi^* \tilde \partial_i \tilde \psi\right)\,.\label{eq6}
\end{alignat}
The total angular momentum in $\tilde x$ , $\tilde y$  and $\tilde z$  direction in term of stress-energy tensor can be defined as \cite{Weinberg:1972},
\begin{alignat}{2}
\tilde J^{ij}=\int(\tilde x^i\,\widetilde T^{j 0}- \tilde x^j\,\widetilde T^{i0})\,d^3 \tilde {x}\,.\label{eq7}
\end{alignat}

For considered initial distribution in equation \eqref{eq5}, the total angular momentum in $\tilde x$ and $\tilde y$ direction ($\tilde J^{2\,3}$ and $\tilde J^{3\,1}$ respectively) are zero. Therefore, one gets the total angular momentum of the system to be $\mathcal{\tilde L_{\rm tot}}=\tilde J^{1\,2}= l \ \tilde{N}\,$. To find the vorticity for the system, we introduce the ``velocity",  ${\bf v}=\bm{\mathcal{J}}/\rho$. Here, $\bm{\mathcal{J}}$ represents the current density and $\rho=|\psi|^2$. Current density $\bm{\mathcal{J}}$ is defined from equation \eqref{eq1} as  $\bm{\mathcal{J}}=i/2 \left(\psi{\bm\nabla} \psi^*-\psi^*{\bm \nabla }\psi\right)$. From these we can now introduce the vorticity ${\bf\Omega}={\bm \nabla}\times{\bf v}$.

In the present work, we have considered the star formation problem when the boson field has a finite self-coupling and angular momentum. 
The presence of attractive self-interaction can also a possibility of forming bound states without gravity \cite{chen:2021} and the presence of angular momentum
may make the star unstable \cite{Dmitriev_2021, Siemonsen_2021}. These processes can introduce new time scales in the problem and
which we shall discuss in the Results and Discussion in more detail.

The star formation process in the presence of self-gravity and self-coupling may lead to an upper bound on  $\tilde{g}$ and it can also modify the estimate of the Jeans length.  If one replaces all spatial derivatives terms from  in GPP equations(i.e. \eqref{eq3}, \eqref{eq4}) with $1/\tilde{L}$, 
where $\tilde{L}$ is the length scale, we get,
\begin{equation}
i \frac{\partial \tilde \psi}{\partial t}\,=\,
\left[  \frac{1}{2 \tilde{L}^4}  - \left(|\tilde{\psi}|^2-\tilde{n}\right) +\tilde{g} \frac{|\tilde{\psi}|^2}{\tilde{L}^2} \right] \tilde \psi \tilde{L}^2 . 
\end{equation}
\noindent
The system would collapse under the gravity.  If there are no particles in the surrounding then a stationary state can be reached i.e. $i \dfrac{\partial \tilde \psi}{\partial t} \sim \omega \tilde \psi$.  By relabelling  $\tilde{L}$ with symbol $\tilde{L}_B$, where the subscript $B$ indicating a bound state  
\begin{equation}
\omega \tilde \psi \,=\,
\left[  \frac{1}{2 \tilde{L}_B^4}  - \left(|\tilde{\psi}|^2-\tilde{n}\right) +\tilde{g} \frac{|\tilde{\psi}|^2}{\tilde{L}_B^2} \right] \psi \tilde{L}_B^2 . 
\end{equation}
Here the term with $\tilde{n}$ is neglected in comparison with $ |\tilde{\psi}|^2$ term. The first and second terms in the square bracket correspond to quantum pressure and self-gravity effects respectively, while the third term is the contribution from the self-interaction.  For a gravitationally bound state to be formed in the collapse, we require that the second term in the square bracket remain larger than the term with self-coupling and thus the following condition to be satisfied,
\begin{equation}
|\tilde{g}\tilde{L}^{-2}_B|\,<\,1
\label{bound}
\end{equation} 
\noindent  
Thus, we can estimate the size of the bound structure by solving the biquadratic equation, \begin{equation}
\tilde{L}^2_B = - \frac{\omega - \tilde{g} |\tilde{\psi}|^2}{2 |\tilde{\psi}|^2} + \sqrt{\frac{(\omega - \tilde{g} |\tilde{\psi}|^2) ^2 + 2 |\tilde{\psi}|^2}{4 |\tilde{\psi}|^4} },\label{eqlb}
\end{equation}
\noindent  where we have ignored the negative square root solution as it could make $\tilde{L}^2_B\,<\,0$. From the above equation, one can see that for attractive self-interaction i.e. $\tilde{g}<0$, the structure formed is smaller in size in comparison with the cases with  $\tilde{g}>0$  and $\tilde{g}=0$. 

\begin{figure*}
	\centering
	\subfloat[] {\includegraphics[width=8cm,height=8.5cm]{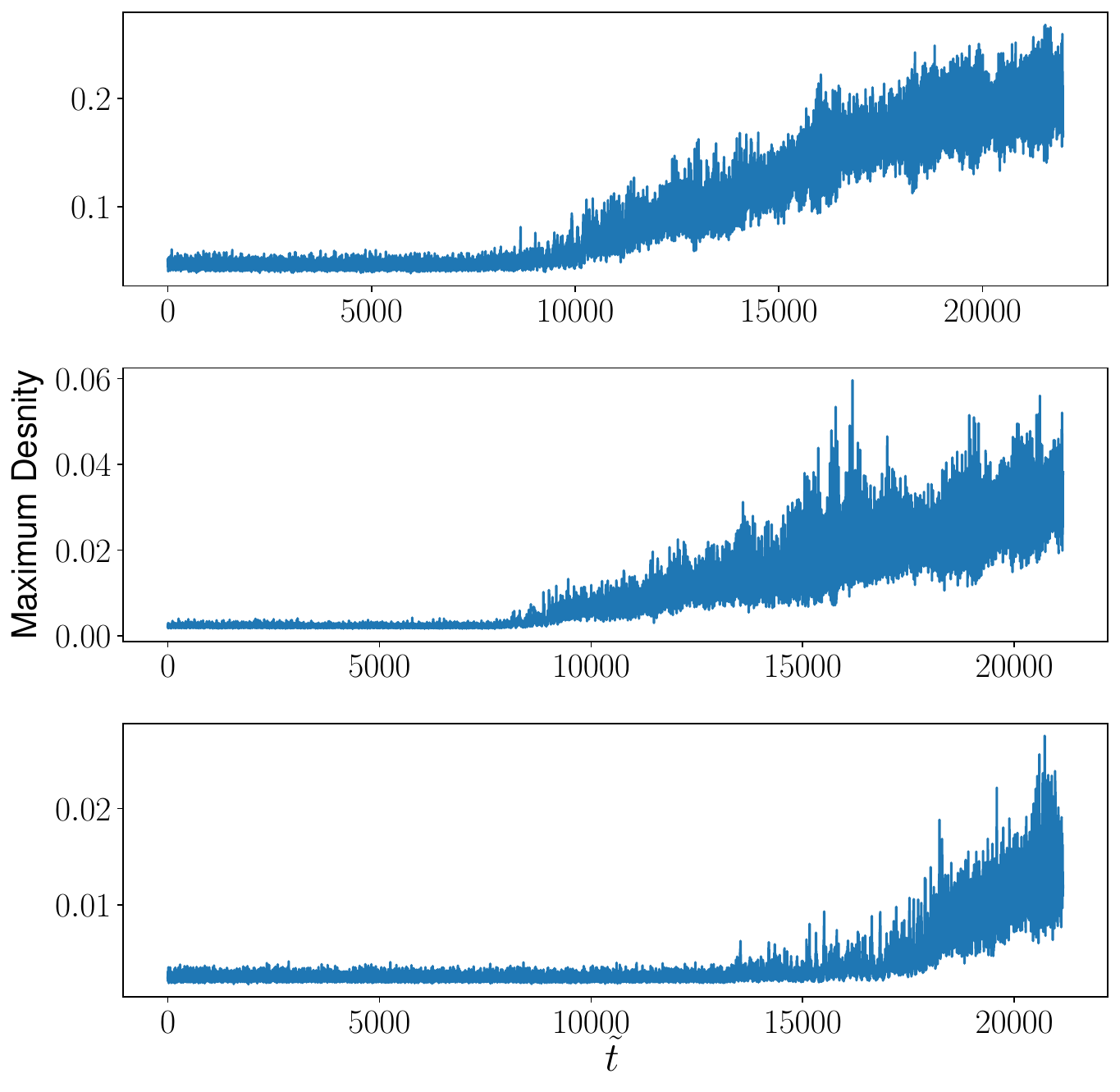}\label{plot4a}\hspace{1cm}}
	\subfloat[] {\includegraphics[width=8cm,height=8.5cm]{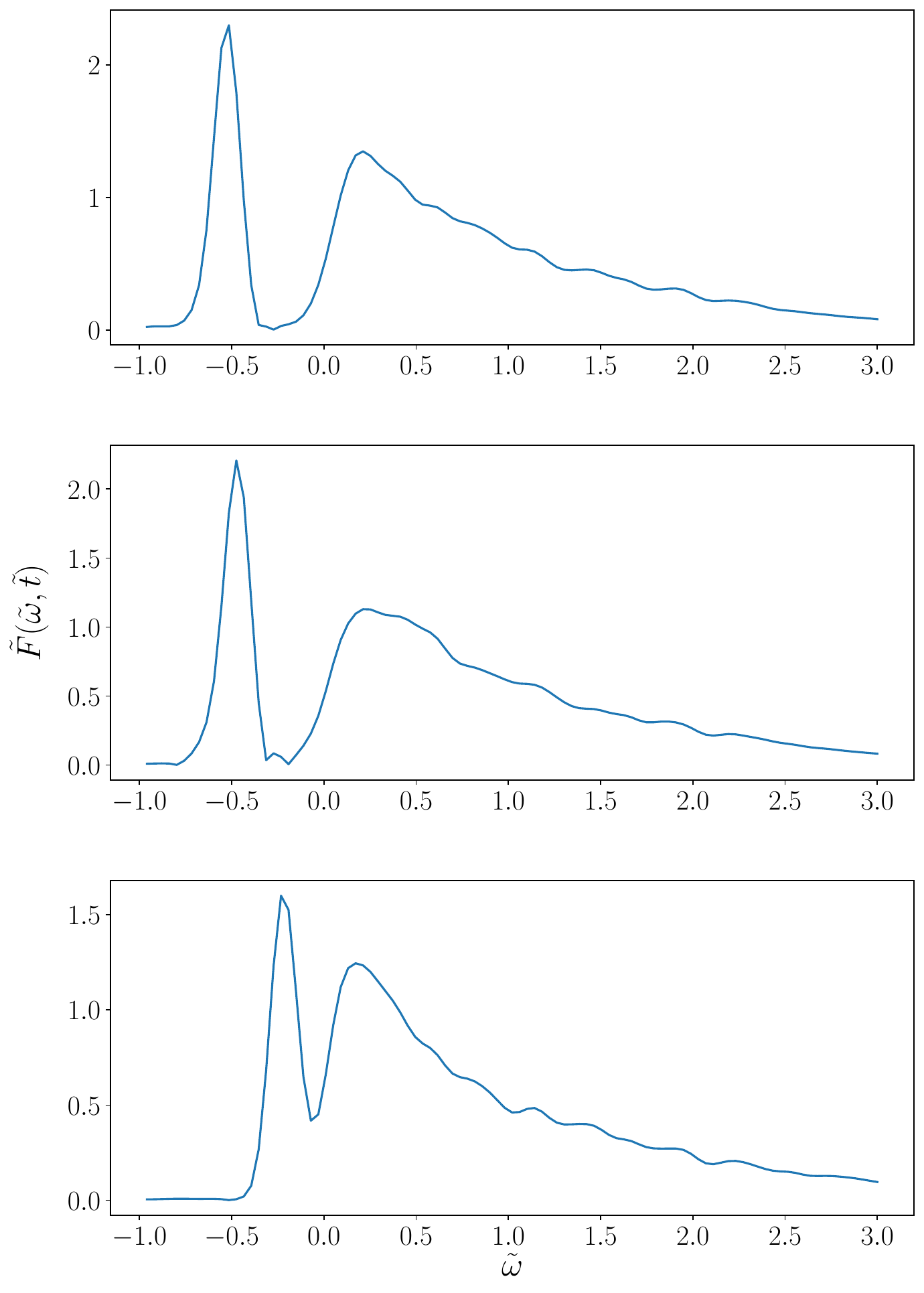}\label{plot4b}}
	\caption{Figure \eqref{plot4a} represents the time evolution of the maximum amplitude of $|\psi|$ for different values of initial angular momentum of bosonic field for $\tilde{g}=0$. Figure \eqref{plot4b} represents the energy spectrum, $\tilde F(\tilde\omega)$, as a function of $\tilde \omega\,$. In both figures, $\mathcal{\tilde L_{\rm tot}}$ varies from top to bottom as 0, 3 and $5$. In figure \eqref{plot4b}, all plots are obtained at $\tilde{t} \sim $ 21000.}
	\label{plot4}
\end{figure*} 
\begin{figure*}
	\includegraphics[width=18cm,height=5.5cm]{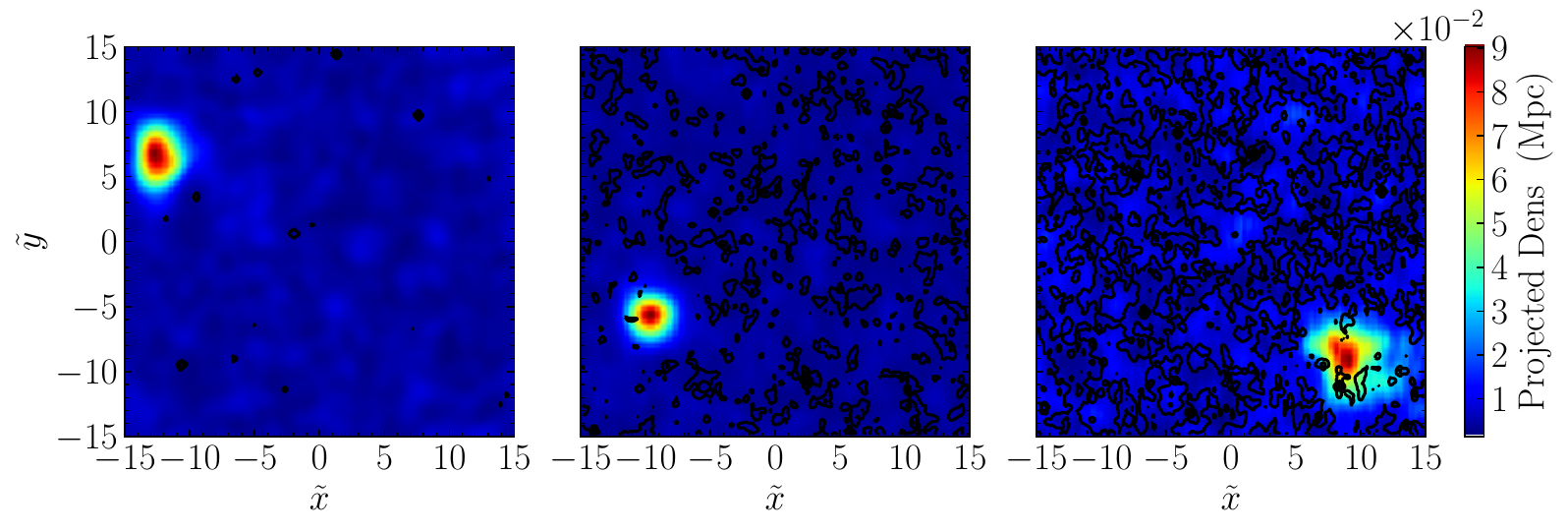}
	\caption{  Three snapshots of density profile $|\tilde{\psi}|^2$ together with magnitude of vorticity  in  $\tilde x\tilde y-$plane are shown at time  $\tilde{t} \sim $ 21000. This the case when there is no self-interaction.
		The first plot on the left corresponds to the case with no angular momentum ($\mathcal{\tilde L_{\rm tot}}=0$), while the second and the the third plots respectively describe the  cases when $\mathcal{\tilde L_{\rm tot}}=3$ and 5. The black contours represent vorticity magnitude.
		The high density region inside the redspots corresponds to the Bose star. The second and third plots shows the presence of vorticity outside the star formation region. In $\tilde x\tilde z-$plane \& $\tilde y\tilde z-$planes also vorticity found to be in the region outside the star.}\label{plot3}
\end{figure*}

\noindent The restriction on values of coupling constant can also be obtained using a different set of arguments \cite{Chen_2022}. Here, the time scale $\tau_{self}$ for obtaining a bound state without a self-gravity is given by $\tau_{self}=\frac{4 \sqrt{2} mv^2}{3 n^2 g^2 \pi}$ when there is an attractive interaction between the constituent particles of the initial cloud. For the case when both self-interaction and self-gravity are present, their combined effect can give rise to a new time scale  $\tau_{eff}$ for structure formation and it is given by, 
 %
\begin{equation}
    \tilde{\tau}_{eff}=\frac{\tilde{\tau}_{gr}\tilde{\tau}_{self}}{\tilde{ \tau}_{gr}+\tilde{\tau}_{self}}.
\end{equation}
\noindent
Thus, when the condition $\tau_{gr}\,<\,\tau_{self}$ is satisfied, self-gravity can play a dominant role in the structure formation. 


\begin{figure*}
	\includegraphics[width=18cm,height=5.5cm]{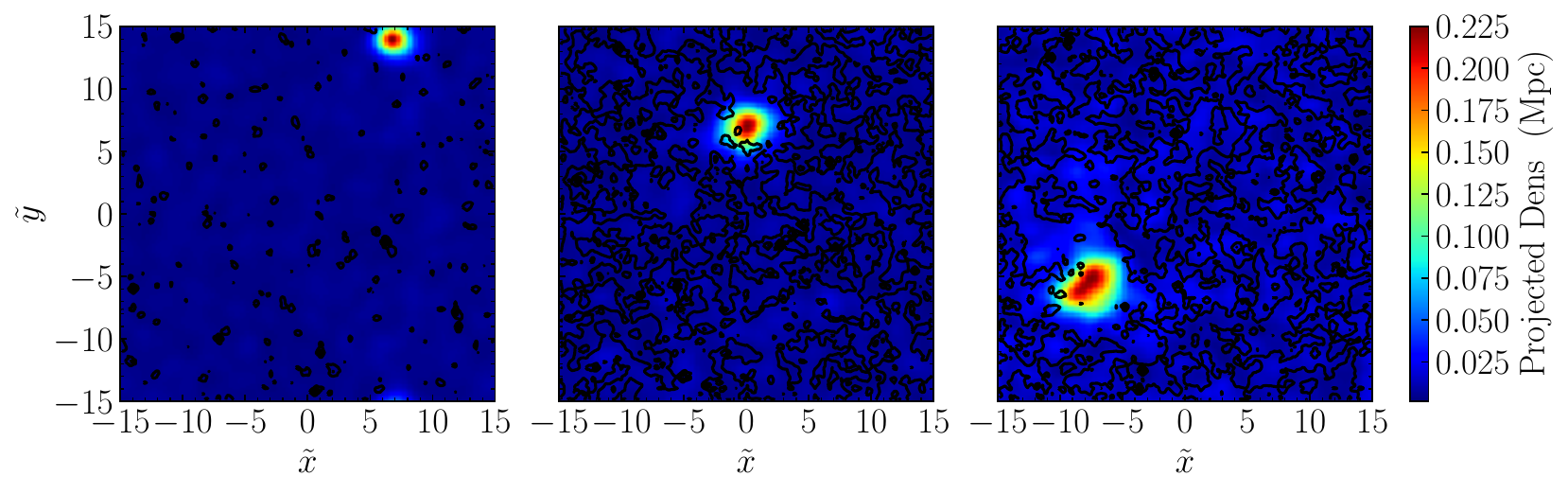}
	\caption{The caption is same as in figure \eqref{plot3}, except here, we have considered attractive self-interaction ($\tilde g=-4.56$) and all plots are obtained at $\tilde t=16000$. }\label{plot5}
\end{figure*}

\begin{figure*}
	\centering
	\subfloat[] {\includegraphics[width=8cm,height=8.5cm]{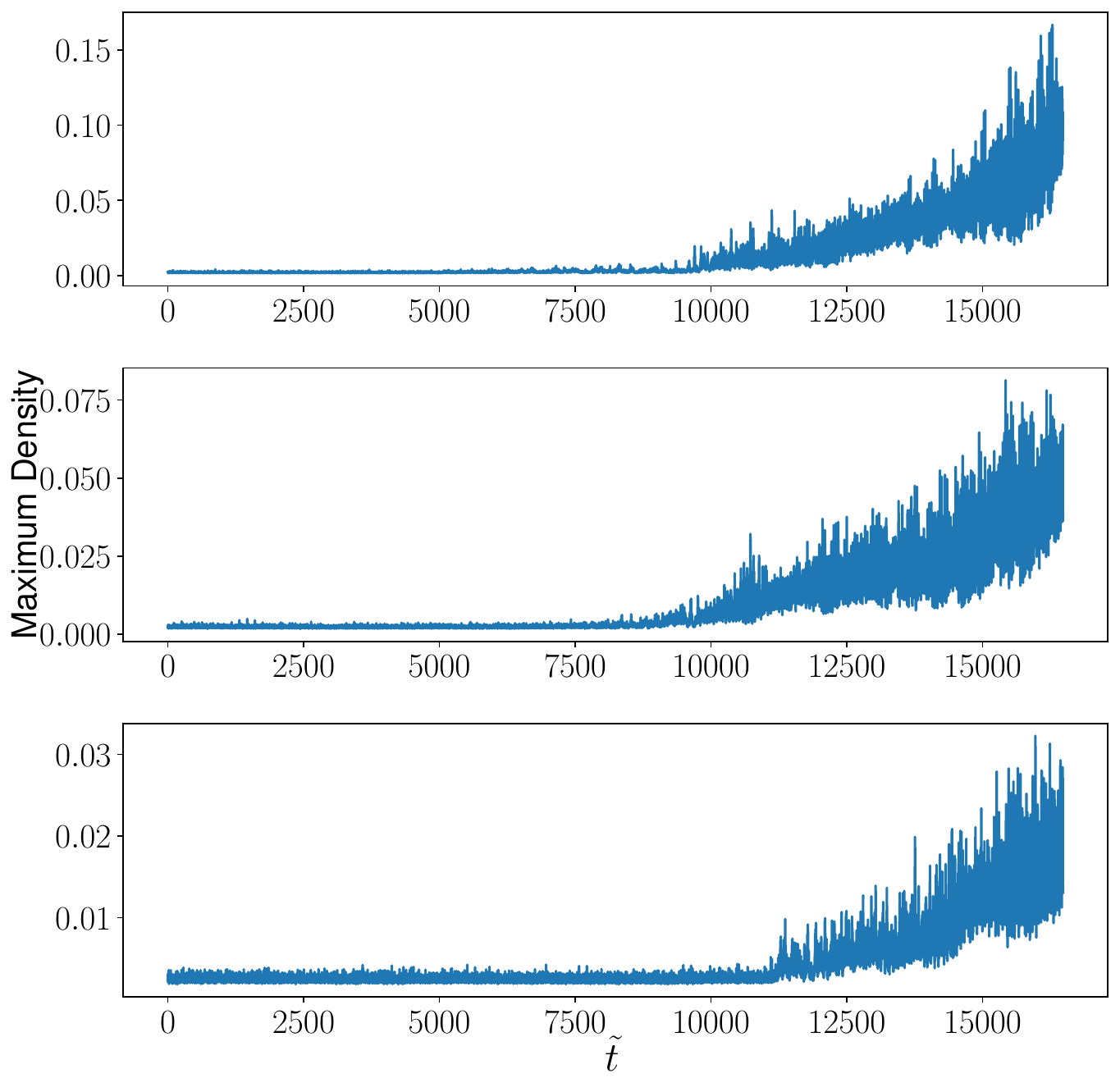}\label{plot6a}\hspace{1cm}}
	\subfloat[] {\includegraphics[width=8cm,height=8.5cm]{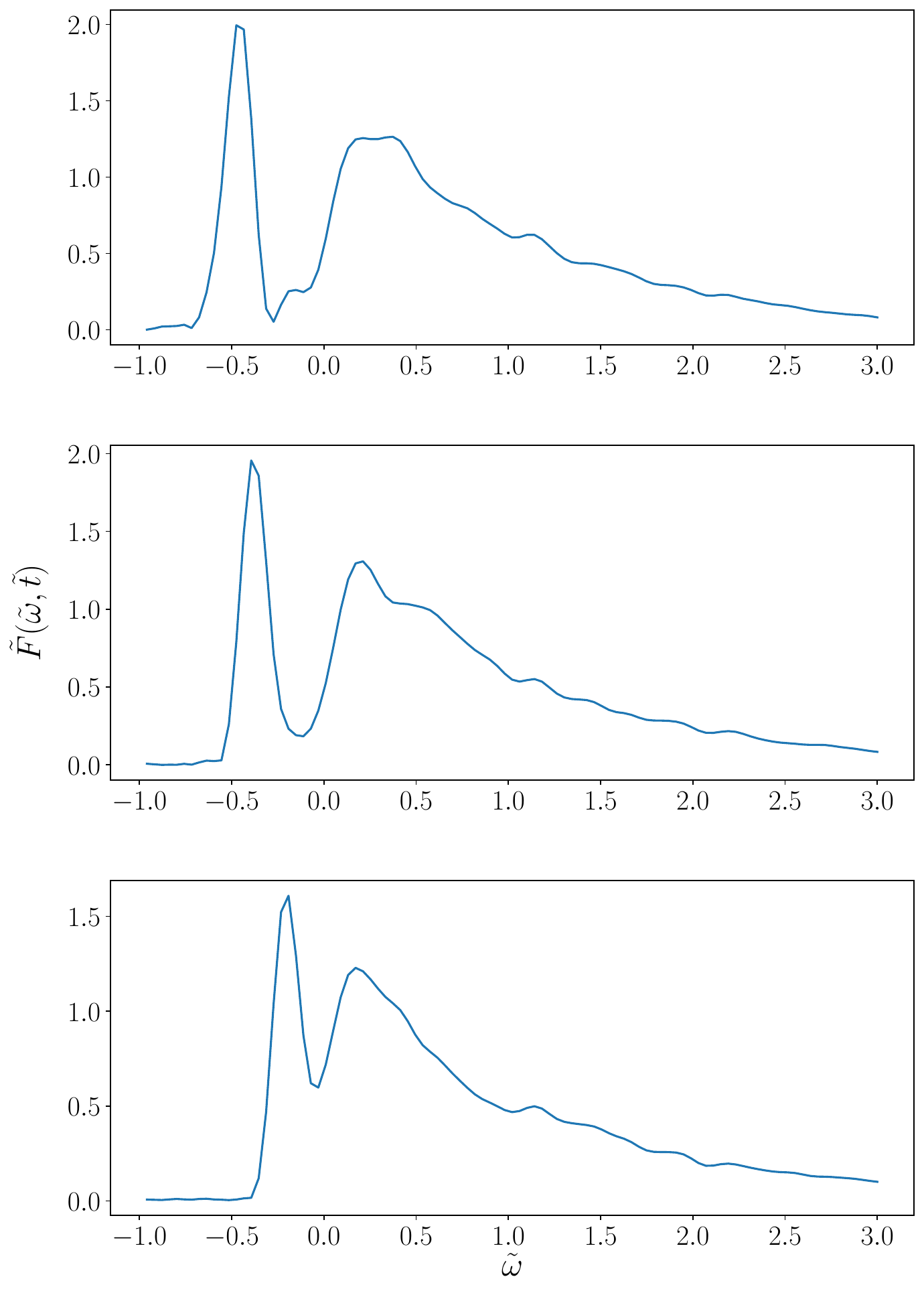}\label{plot6b}}
	\caption{The caption is same as in figure \eqref{plot4}, except here, we have considered attractive self-interaction ($\tilde g=-4.56$) and all plots are obtained at $\tilde t=16000$. }
	\label{plot6}
\end{figure*}  
\section{Results and Discussion}\label{sec3}


 The density profiles of the initial wave-functions in $yz,\,zx\,\&\,xy$ planes for various values of angular-momentum $\mathcal{L}_{tot}=0$ are shown in figure(1).
 In absence of angular momentum or self-coupling in the initial boson cloud, the only important time scale is the gravitational condensation time described by the formula given in Ref. \cite{Levkov_2018},
\begin{equation}
\tau_{gr} = \frac{b \sqrt{2}}{12 \pi^3} \frac{mv_0^6}{ G^2 { n}^2 \Lambda},
\end{equation}\label{timeself}
\noindent where,   $\textit{n}$ is the boson number density and , $\Lambda$ = $\log(mv_0\,L)$ with   $\tilde{L}\,=\,mv_0 L$ being the box length used in the numerical simulation. It should be noted that the complete determination of parameter $b$ is not possible but can be a number of the order of unity\cite{Levkov_2018}. Using the dimensionless units we write the gravitational condensation time as:

\begin{equation}
\tilde{\tau}_{gr} = \frac{b \sqrt{2}}{12 \pi^3} \frac{1}{\tilde{n} \,log(\tilde{L})}.
\end{equation}\label{DLtimeself}

Next, keeping the available computational resources in mind,   we choose $\tilde{N}=10$, $ \tilde {L}= 30$,  and resolution  $\tilde{L}^3/128^3$. This would imply that the estimate of $\tilde{\tau}_{gr} \sim 8\times 10^3$.  The continuum limit with the above resolution has been discussed in \cite{Levkov_2018, chen:2021}.

  To ensure the validity of the numerical results presented here we have done the following checks: In absence of angular momentum and any self-interaction in the scalar field, the numerical code reproduces the results presented by \cite{Levkov_2018, Schwabe_2020}. These checks include that we get similar values of $\tilde{\tau}_{gr}$ and having a spike in the power spectrum of the wave-function $F(\omega, t)$ (defined below) at negative values of frequency when time $\tilde{t}$ satisfies the condition  $\tilde{t}\,>\,\tilde{\tau}_{gr}$. 
 Another test is based on the conservation of total energy and it is satisfied by all the numerical results presented in this manuscript. Here, we calculate the total energy of the system for the initial data and compare it with all the subsequent steps of time. If for any time step the total energy differs by more than 5 percent, we do not run the code any further.

For given values of $\tilde{N}=10$ and $\tilde{L}=10$, the highest allowed value of $|\tilde{g}|$ is 4.64 for the formation of a gravitationally bound state[\eqref{bound}].   For $|\tilde{g}|\,>\,4.64$ no star formation is seen during the time scales $\tilde{t}$ ($>\tilde{\tau_{gr}})$ allowed by the convergence test imposed on our program. For the case when the repulsive interaction with $\tilde{g}.4.64$ is considered, we can not see any star formation happening during the time interval allowed by our conservation test. Further, we would like to note that if one reduces the value of $\tilde{g}$ by an order of magnitude, then the results obtained are very similar to the case with $\tilde{g}=0$. Thus in this work, we present the results for the following three values of the coupling constant $\tilde{g}=-4.64\ ,0.0\ , \&  +4.64$.
The data generated by the code is analysed by using \texttt{The yt Project} \footnote{\href{https://yt-project.org/doc}{https://yt-project.org/doc}} \cite{Turk_2010}.

  Here we would like to note that recently in \cite{Levkov_2018, Siemonsen_2021} the authors have studied the stability of an isolated rotating Bose star. It is demonstrated that when the self-interaction among the constituent particles is attractive or negligibly small, the star becomes unstable for {\it any value of angular momentum}. The instability results in the shedding of
 the particles (also the angular momentum)  in the region outside the star. The instability results in forming of a Saturn ring-like structure around the star.
 The time scale associated with the instability of a rotating boson star \cite{Dmitriev_2021} was estimated using the growth rate given by
\begin{equation} \tau_{ins} = \frac{100}{2.2} \frac{l^2}{\alpha_l} \frac{1}{m^3 G^2 M_s^2}, 
\end{equation}  
\noindent where, $l,\, M_s$ ,and $\alpha_l$  represent the values of angular momentum, the mass of the star ,and an angular-momentum dependent parameter, respectively.
 However, the authors demonstrate that when the self-interaction between the constitutive particles is repulsive, it is possible to have a stable Bose star configuration. 
 
Thus it would be rather interesting to see whether Bose stars formed in our numerical results are rotating or not. First, we would like to emphasize that 
  in the above stability analysis \cite{Dmitriev_2021} the initial configuration is assumed to be 
  an isolated boson star. In our work, the star formation occurs after time $\tilde{\tau}_{gr}$ from the initial state with a randomly distributed density over the entire space.  To understand how the instability influences
  the presented numerical results, first, consider the ratio  ${\tilde{\tau}_{ins}}/{\tilde{\tau}_{gr}}$.
   Using the estimates for a typical particle velocity from \cite{Levkov_2018}  $v_0\sim 30 $~km/s together with $\tilde{n}\,\&\, \tilde{L}$ , the ratio ${\tau_{ins}}/{\tau_{gr}}\sim 10^{-2}$. Further, it is to be noted that we have used the value of $\tilde{\tau}_{gr} \sim 8000$. The typical range of $\tilde{\tau}_{gr} \sim 8000-16000$ for $\tilde{g}\geq\,0$. Thus one can only decrease  ${\tilde{\tau}_{ins}}/{\tilde{\tau}_{gr}}$ by increasing  $\tilde{\tau}_{gr}$. Since ${\tau_{ins}}/{\tau_{gr}}\leq 10^{-2}$, we believe that the possible instability would have time to saturate.

 To check that when the star condensation begins we employ the following three tests: 1)
The first one is to study the time evolution of the maximum amplitude $|\tilde{\psi}(\tilde x, \tilde t)|$. The typical value for the initial maximum amplitude $|\tilde{\psi}(\tilde x, \tilde t)| \sim 2\times10^{-2}$, the sudden rise in the value of maximum amplitude could be either due to gravity or self-interaction produced bound state. This provide us the estimate of 
$\tilde{\tau_{eff}}$ for $\tilde{g}\leq 0$ or $\tilde{\tau_{gr}}$ when 
$\tilde{g}\,>0$ for a repulsive interaction. 
2)  The second test is about studying the power spectrum $F(t,\omega)$ of $\psi(x,t)$ \cite{Levkov_2018}, defined as a Fourier image of the correlator,
\begin{equation}
F(\omega, t) = \int \frac{dt_1}{2 \pi} d^3\bm{x}\  \psi^*(t,\bm {x})\, \psi(t+t_1,\bm{x})\, e^{i \omega t_1 - t_1^2/\tau_1^2}.
\end{equation}\label{spectrum}
Around the time when the gravitational condensation happens, $F(\omega, t)$  starts developing a peak at $\omega_s < 0$ which indicates the formation of a gravitationally bound state. 
3) One can also check if the formed structure is gravitationally bound, then it should satisfy the condition $|\tilde{g}\tilde{L}^{-2}_B|\,<\,1$.

\begin{figure*}
	\includegraphics[width=18cm,height=5.5cm]{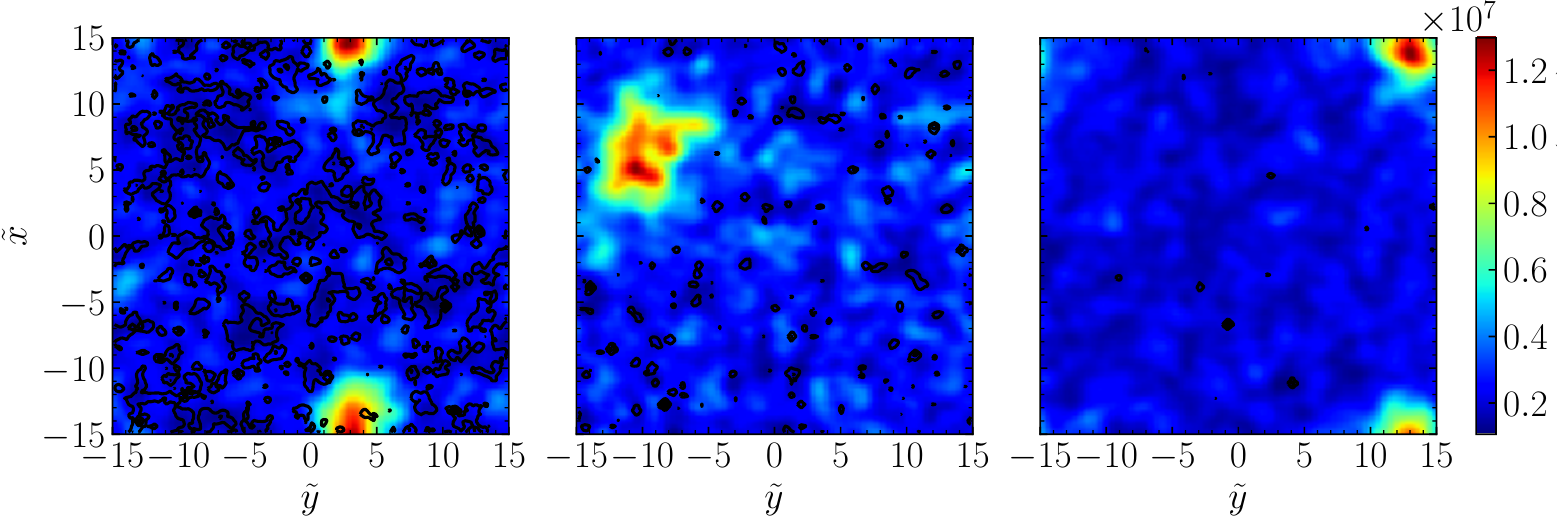}
	\caption{Vorticity magnitude plotted in $\tilde{x}\tilde{y}$ plane at times $\tilde{t} = 21972.6, 23345.9 \textrm{and} 27465.7$ from left to right together with the Density for $\tilde g=+4.56$ (repulsive self-interaction) and $\mathcal{\tilde L_{\rm tot}}\,=\,5.0$. With the passage of time the net vorticity vanishes from the surrounding of the star formation region. The plots does not have any vorticity outside the star forming region. The absence of vorticity is seen in the other two planes also. This may be indicative of angular momentum entering the star. }\label{plot7}
\end{figure*}

\begin{figure*}
	\centering
	\subfloat[] {\includegraphics[width=8cm,height=8.5cm]{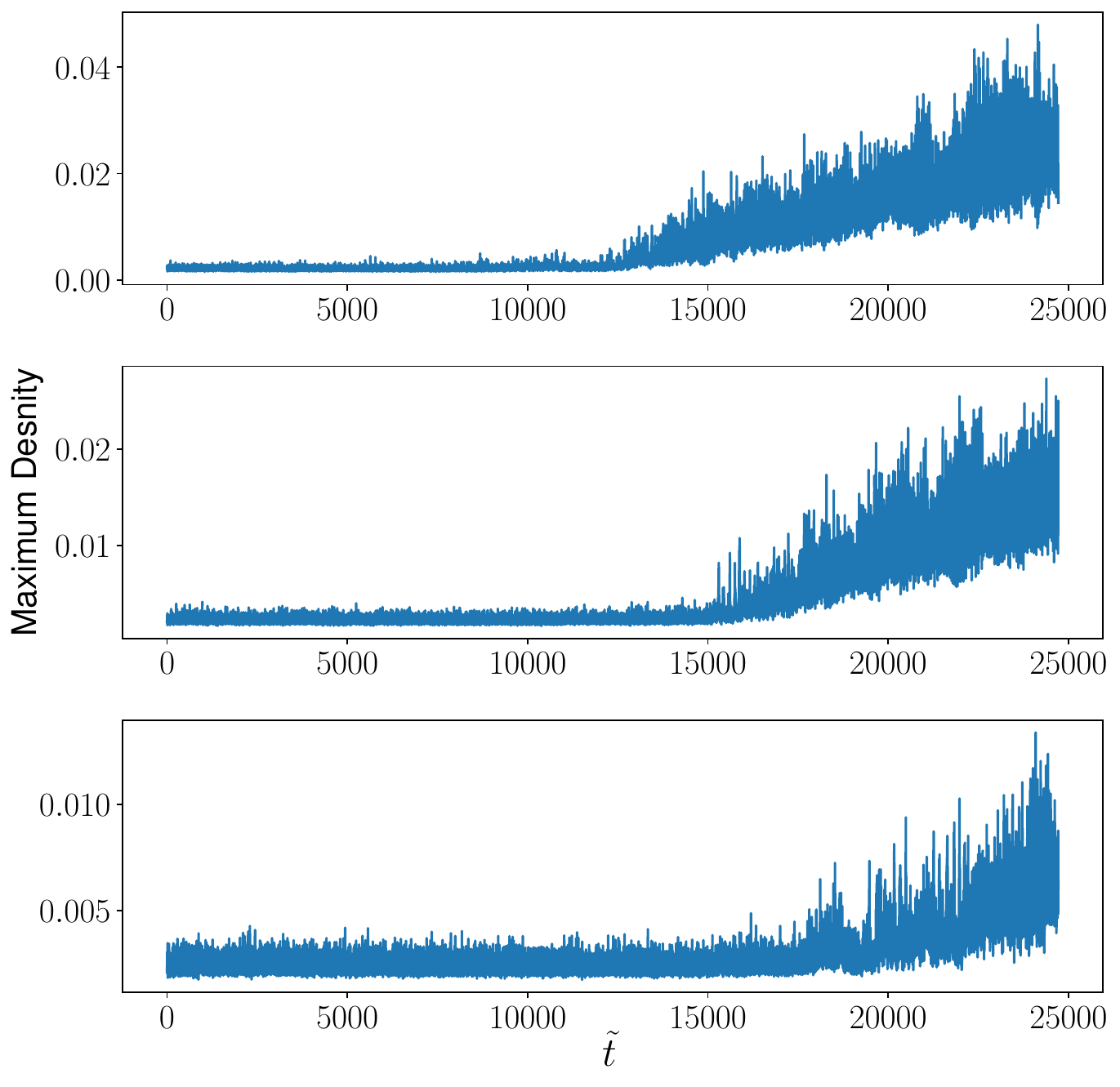}\label{plot8a}\hspace{1cm}}
	\subfloat[] {\includegraphics[width=8cm,height=8.5cm]{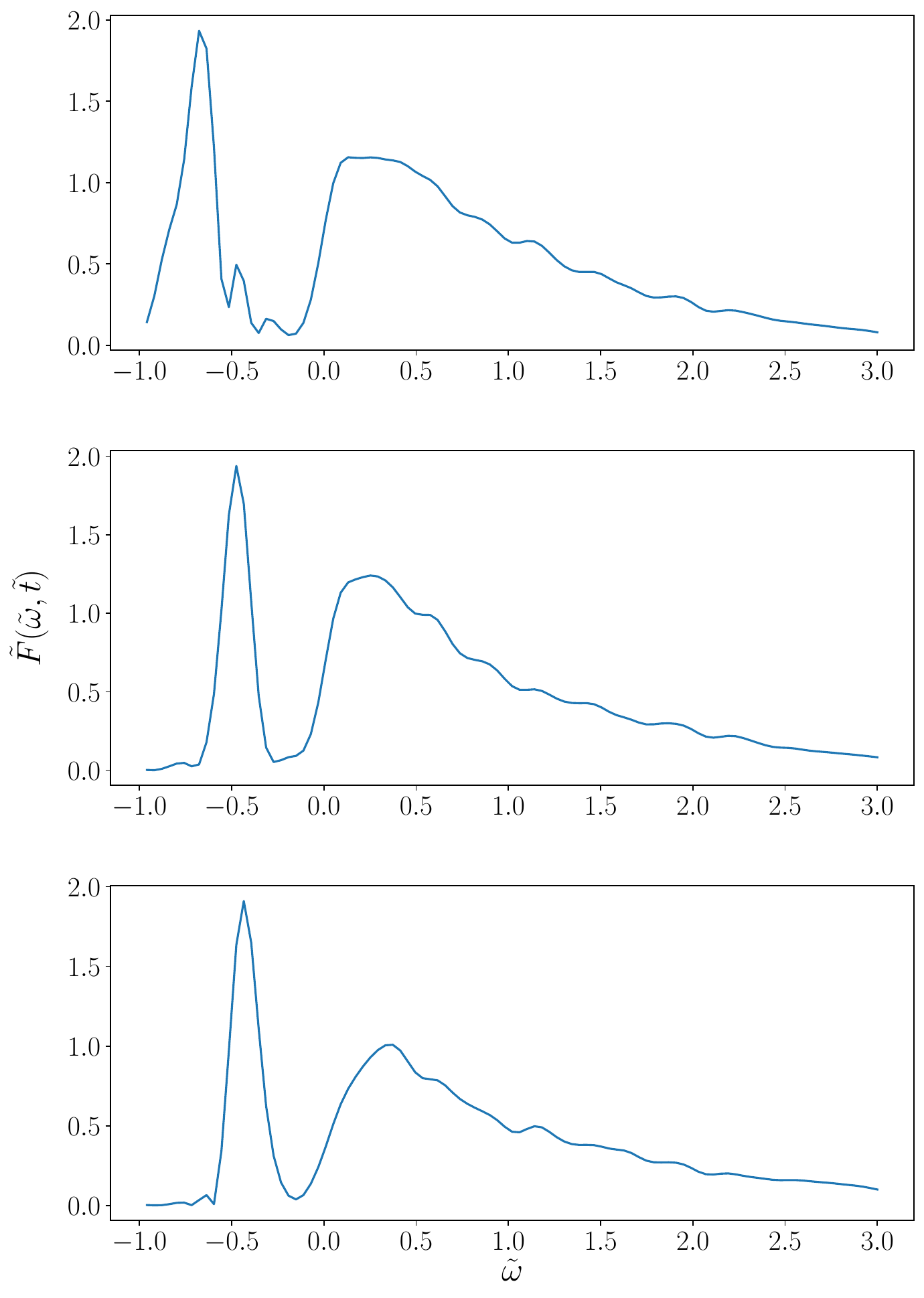}\label{plot8b}}
	\caption{The caption is same as in figure \eqref{plot4}, except here, we have considered repulsive self-interaction {($\tilde g=4.56$)} and all plots are obtained at $\tilde t=25500$..}
	\label{plot8}
\end{figure*}  
\subsection{Bose star formation  without any self-interaction}

Figure \eqref{plot4a} shows the evolution of maximum density  $\max|\tilde\psi|^2$ with time $\tilde{t}$. The plots from the top to bottom respectively correspond to cases with  $\mathcal{\tilde L_{\rm tot}} =0,\, 3$ and $5$.   $\max |\tilde{\psi}|$ remains constant and small for a long time $\tilde{t}\sim\tilde{\tau}_{gr}$ and then there is a sudden rise in its value. It is to be noted that the mean value of density continues to rise monotonically for $\tilde{t}\,>\,\tilde{\tau}_{gr}$  till 
the convergence test stop the code around time $\tilde{t}\sim$~ 21000. 
Ideally, the growth should saturate at some point in time. But due to the restriction imposed by the convergence test,
  it is not possible to check this in our numerical results. The density saturation after the formation of a Bose star cloud not be reached in the earlier studies for $\mathcal{\tilde L_{\rm tot}}=0$ in \cite{Levkov_2018, chen:2021} due to the computational limitations.
Table(I) provides estimates of $\tilde{\tau}_{gr}$ for different values of  $\mathcal{\tilde L_{\rm tot}}$.  The table shows that if one increases angular momentum value from 0 to 3, the value of $\tilde{\tau}_{gr}$ decreases. This feature is counterintuitive, as one would expect that increasing angular momentum in the collapsing would result in increasing the condensation time. As we shall see later, similar  behavior of $\tilde{\tau}_{gr}$  is also seen in the cases for the case of attractive interaction ( $\tilde{g}\,<\,0$).
However, if one increases  $\mathcal{\tilde L_{\rm tot}}$ from 0 to 5, there is a significant increase in $\tilde{\tau}_{gr}$. 

   Figure\eqref{plot4b} describes the behaviour of spectrum $F(\omega, t)$ as a function of frequency at $\tilde{t}\sim 21000$, which larger than the condensation time $\tilde{\tau}_{gr}$ for all the three cases described by Table(I). The figure clearly shows the presence of a prominent minimum
 in the region where the frequency has negative values for all the different values of   $\mathcal{\tilde L}_{\rm tot}$. 
  Here we would like to emphasise  that the minimum in $F(\omega, t)$ occurs soon after $\tilde{\tau}_{gr}$ and it continues to persists all the time till
 the convergence test imposed on the code remains valid. The existence of the peak in the negative frequency regime implies the existence of a gravitationally bound state or a Bose star.

To illustrate how the density of the self-gravitating cloud evolve with time, figure(3) shows snapshots of density at different values of time when the initial cloud has angular momentum $\tilde{\mathcal{L}}_{tot}=5$.
 The high-density regions are shown with red color. At the time $\tilde{t}\sim \tau_{gr}$ all the high-density regions come close to each other and get localized in some region of the box. The density in this region increases with time. Once the star condensation begins at $\tilde{t}\sim\tilde{\tau}_{gr}$ it persists till the time the convergence test stops the program
 
 After the gravitational condensation, the typical time interval during which Bose-star is seen the numerical result is around 0.5$\tilde{\tau}_{gr}$$\,-\,1.5\tilde{\tau}_{gr}$, and value of $\tilde{\tau}_{gr}$ depends on $\mathcal{\tilde L}_{\rm tot}$. As we have discussed earlier, a rotating  boson star could be unstable and the instability time scale is much shorter than the condensation time i.e.  $\frac{\tilde{\tau}_{ins}}{\tilde{\tau}_{gr}}\sim 10^{-2}$,
 Since the negative peak in $F(\omega, t)$ is seen for much longer than time  ${\tilde{\tau}_{ins}}$ in our simulation, one can argue that the Bose stars seen in our numerical results are stable. 
 Therefore, it is interesting to see what happens to the angular momentum present in the initial cloud. To get some insight into this problem, in figure(3) we have shown three snapshots of density profile $|\psi|^2$ around time 2100 along with the vorticity magnitude in $\tilde{x}\tilde{y}$-plane. The vorticity magnitude is shown by the black contours. The leftmost plot in figure(2) shows the case with $\mathcal{\tilde L}_{\rm tot}=0$, here there is no initial vorticity. But in the other two plots $\mathcal{\tilde L}_{\rm tot}\neq\,0$,  all the black contours are located outside the star formation region. In other two planes also similar situation is seen.
  Thus we believe that the Bose star for $\tilde{g}=0$ may not have any intrinsic angular momentum and our results are consistent with the stability
  analysis of  \cite{Dmitriev_2021}.
 
 Table \eqref{tab 2}, lists the values of central density, $R_{95}$ of the star and mass of the star at time $\tilde{t}\sim 2100$ for different values of $\mathcal{\tilde L}_{\rm tot}$.  $R_{95}$ measures distance from the centre at which 95\% density of the star is located. 
 The total mass of the star is calculated for a spherical volume enclosed in $R_{95}$ for different values of $\mathcal{\tilde L}_{\rm tot}$.
  In the non-relativistic limit, a self-gravitating cloud of bosons can clump at a length scale larger than the Jeans length\cite{chen:2021}.
  However, in this work quantity,  $L_B$ is given by equation \eqref{eqlb} which plays a role similar to the Jeans length.
   One can estimate $L_B$ by using the values of Central density from Table (II) and estimate of $\omega$ from corresponding plots of the spectrum. This gives an estimate of $L_B\sim 2.0$  for $\mathcal{\tilde L}_{\rm tot}=0$. For this case f $R_{95}$ is 4.48 and we have $R_{95}\,>\,L_B$.

   It is found in Ref. \cite{Levkov_2018}, in absence of angular-momentum in the initial cloud,
   that  the Bose-star begins to condense in a rarified and non-compact state, but as its mass increases it becomes more compact. But this assertion changes when the initial collapsing cloud has non-zero angular momentum: Consider the two cases with  $\mathcal{\tilde L}_{\rm tot}\,=$\, 0 \& 3  in Table(II).  For $\mathcal{\tilde L}_{\rm tot}\,=3$ case $\tilde{\tau}_{gr}\sim 8500$ and thus it begins to form earlier than the star with $\tilde{\tau}_{gr}\sim 9500$. But the star formed earlier has less mass and is more compact than the star formed later. One can understand when the initial cloud has $\mathcal{\tilde L}_{\rm tot}\,= 3$, it hindered the mass growth of the star and as a result, it remains less massive as the star in the case with $\mathcal{\tilde L}_{\rm tot}\,=0$. But the less massive star is 
   the most compact one in Table (II). For the case, $\mathcal{\tilde L}_{\rm tot}\,=5$, the star has the smallest mass and central density while it is the least compact. The tabulated values of the central density and $R_{95}$ show the trend that more the central density implies less values of $R_{95}$(compactness) when the initial collapsing cloud has net angular momentum.

\subsection{Bose star formation in  presence of self-interaction} 

  Next, we study Bose star formation in presence of the finite self-interaction.  We studied the gravitational collapse of the self-interacting boson field for various values of the coupling constant $g$ and as discussed above that for an arbitrarily large value of the coupling strength star formation may not be possible. For the values of  $\tilde{N}=10$ and $\tilde{L}=30$  we found that maximum allowed  value of $|\tilde{g}|\sim 4.56$. However, if we reduce the values of the coupling constant by an order of magnitude then the results approach the case with no self-interaction. Therefore, in this section we shall restrict our analysis to a very high value of the coupling.

\begin{table}[h]
	\caption{\label{tab1} Gravitational Condensation Time ($\tilde{\tau}_{gr}$) \& Amplitude growth rate ($c$) for No Self-Interaction. All the numbers are dimensionless in the rescaled units as defined in the main text.}
	\begin{ruledtabular}
		\begin{tabular}{cccccccc}
			& $\tilde{\tau}_{gr}$\\
			\hline
			\\
			\vspace*{0.1cm}
			$\mathcal{\tilde L_{\rm tot}}=0.0 $ & 9500   \\
			\vspace*{0.1cm}
			$\mathcal{\tilde L_{\rm tot}}=3.0$ & 8500    \\
			\vspace*{0.1cm}
			$\mathcal{\tilde L_{\rm tot}}=5.0$  & 16200  \\
		\end{tabular}
	\end{ruledtabular}
\end{table}
\begin{table}[h]
	\caption{\label{tab 2} Central density, radius \& mass corresponding to Bose star at time $\tilde{t} \simeq 21000\,$ for No Self-Interaction. All the numbers are dimensionless in the rescaled units as defined in the main text.}
	\begin{ruledtabular}
		\begin{tabular}{cccccccc}
			& Central Density & $R_{95}$ & Mass\\
			\hline
			\\
			\vspace*{0.1cm}
			$\mathcal{\tilde L_{\rm tot}}=0.0 $   & 0.028    &4.15 &1.02 \\
			\vspace*{0.1cm}
			$\mathcal{\tilde L_{\rm tot}}=3.0$&    0.041  & 2.92 &0.85   \\
			\vspace*{0.1cm}
			$\mathcal{\tilde L_{\rm tot}}=5.0$ & 0.012  & 4.32 &0.67 \\
		\end{tabular}
	\end{ruledtabular}
\end{table}

\begin{table}[h]
	\caption{\label{tab3} Gravitational Condensation Time ($\tilde{\tau}_{gr}$) \& Amplitude growth rate ($c$) for Attractive Self-Interaction. All the numbers are dimensionless in the rescaled units as defined in the main text.}
	\begin{ruledtabular}
		\begin{tabular}{cccccccc}
			& $\tilde{\tau}_{gr}$ & $c$ \\
			\hline
			\\
			\vspace*{0.1cm}
			$\mathcal{\tilde L_{\rm tot}}=0.0 $   & 9200   &0.55  \\
			\vspace*{0.1cm}
			$\mathcal{\tilde L_{\rm tot}}=3.0$& 9200    & 0.48    \\
			\vspace*{0.1cm}
			$\mathcal{\tilde L_{\rm tot}}=5.0$ & 11000  & 0.43  \\
		\end{tabular}
	\end{ruledtabular}
\end{table}

\begin{table}[h]
	\caption{\label{tab4} Central density, radius \& mass corresponding to Bose star at time $\tilde{t} \simeq 16500\,$ for Attractive Self-Interaction. All the numbers are dimensionless in the rescaled units as defined in the main text.}
	\begin{ruledtabular}
		\begin{tabular}{cccccccc}
			& Central Density & $R_{95}$ & Mass\\
			\hline
			\\
			\vspace*{0.1cm}
			$\mathcal{\tilde L_{\rm tot}}=0.0 $   & 0.098    &2.18 &0.85 \\
			\vspace*{0.1cm}
			$\mathcal{\tilde L_{\rm tot}}=3.0$&    0.037  & 2.99 &0.84   \\
			\vspace*{0.1cm}
			$\mathcal{\tilde L_{\rm tot}}=5.0$ & 0.012  & 4.31 &0.73 \\
		\end{tabular}
	\end{ruledtabular}
\end{table}

\begin{table}[h]
	\caption{\label{tabV} Gravitational Condensation Time (\& $\tilde{\tau}_{gr}$) \& Amplitude growth rate ($c$) for Repulsive Self-Interaction. All the numbers are dimensionless in the rescaled units as defined in the main text .}
	\begin{ruledtabular}
		\begin{tabular}{cccccccc}
			& $\tilde{\tau}_{gr}$  \\
			\hline
			\\
			\vspace*{0.1cm}
			$\mathcal{\tilde L_{\rm tot}}=0.0 $   & 12000     \\
			\vspace*{0.1cm}
			$\mathcal{\tilde L_{\rm tot}}=3.0$& 15000    \\
			\vspace*{0.1cm}
			$\mathcal{\tilde L_{\rm tot}}=5.0$ & 18000  \\
		\end{tabular}
	\end{ruledtabular}
\end{table}
\begin{table}[h]
	\caption{\label{tab VI} Central density, radius \& mass corresponding to Bose star at time $\tilde{t} \simeq 26500\,$ for Repulsive Self-Interaction. All the numbers are dimensionless in the rescaled units as defined in the main text.}
	\begin{ruledtabular}
		\begin{tabular}{cccccccc}
			& Central Density & $R_{95}$ & Mass\\
			\hline
			\\
			\vspace*{0.1cm}
			$\mathcal{\tilde L_{\rm tot}}=0.0 $   & 0.014    &4.48 &0.97 \\
			\vspace*{0.1cm}
			$\mathcal{\tilde L_{\rm tot}}=3.0$&    0.025  & 3.32 &0.75   \\
			\vspace*{0.1cm}
			$\mathcal{\tilde L_{\rm tot}}=5.0$ & 0.028 & 3.25 &0.74 \\
		\end{tabular}
	\end{ruledtabular}
\end{table}
	
\subsubsection{Attractive self-interaction}

  In this case, it may be possible to form bound structures entirely due to the balance between the attractive force provided by the self-interaction and the quantum pressure. Therefore it is necessary to check whether self-gravity plays a significant role in the bound state observed in the numerical results. In this case, we shall consider $\tilde{g}=-4.56$.

  We first analyse the plots of maximum density $\max |\tilde{\psi}|^2$ vs. time for different values of $\mathcal{\tilde{L}}_{\rm tot}$ are shown in figure(5a). All the plots show that for a long time $\max |\tilde{\psi}|^2$ remains very small and constant. After the certain time when the condensation happens there the value of $\max |\tilde{\psi}|^2$ rises monotonically till the time allowed by the convergence test related with the total energy. The specific value of time at which $\max |\tilde{\psi}|^2$ starts rising depends on value of $\mathcal{\tilde{L}}_{\rm tot}$ present in the initial cloud.
  Table( \eqref{tab3}) gives a list estimates of $\tilde{\tau}_{gr}$ for different value of $\mathcal{\tilde{L}}_{\rm tot}$.  It is interesting to note when $\mathcal{\tilde{L}}_{\rm tot}$ is increased from 0 to 3, the value of condensation time does not change. This is similar to the counter-intuitive behaviour of $\tilde{\tau}_{gr}$ we have seen in the case $\tilde{g}=0$ before. 
  But, if value  $\mathcal{\tilde{L}}_{\rm tot}$ changes from 3 to 5, there is a significant increase in the value of $\tilde{\tau}_{gr}$. Since the attractive interaction among the particles of the collapsing cloud is working together with the self-gravity may expect to see a reduction in values of $\tilde{\tau}_{gr}$ in comparison to the case with $\tilde{g}=0$.  The listed values of  $\tilde{\tau}_{gr}$ in Tables(\eqref{tab1}) \& (\eqref{tab3}) confirms this expectation. However, values of central density and $R_{95}$ imply that higher values of central density imply smaller values of $R_{95}$.

  Next, figure \eqref{plot6b} depicts the plots of spectrum $\tilde{F}(\tilde{t},\tilde{\omega})$ versus frequency $\tilde{\omega}$ at  time $\tilde{t}\simeq\, 16000$ for three different values of $\mathcal{\tilde{L}}_{\rm tot}$. All three plots in figure \eqref{plot6b} show the formation of a gravitationally bound state indicated by the presence of a peak in the negative frequency regime. As we have discussed earlier, when the interaction is attractive, there is an upper bound on the coupling constant $\tilde{g}$ given by the condition \eqref{bound}. This requires an estimation of $L_B$. 
  Table (\eqref{tab4}) provides  values of central density, $R_{95}$ \& mass of the Bose star at time 16500 when $\tilde{g}=-4.56$ for the various values of $\mathcal{\tilde{L}}_{\rm tot}$. Using the values of density from Table(\eqref{tab4}) and frequency from \eqref{plot6b}, one finds $L^2_B\sim 5.39$. This means
  that  $ \frac{|g|}{\tilde{L}^2_B} \sim 0.86 < 1 $ i.e. the bound is indeed satisfied. Further, it is to be noted that Table(\eqref{tab3}) shows that for stars for both cases corresponding to  $\mathcal{\tilde{L}}_{\rm tot}\,=0\,\&\,3$ cases, started to condense around the same at the same time. Thus according to Table(\eqref{tab4}), both the stars had the same time growth. Interestingly the star in the case of  $\mathcal{\tilde{L}}_{\rm tot}=0 $ case is more compact in comparison with the other star in spite of the fact that both the stars are having approximately the same masses. Values of central density and $R_{95}$ given in Table(\eqref{tab4}) imply that higher values of central density correspond to smaller values of $R_{95}$. The similar trend between the central density and $R_{95}$ we have seen in the case when $\tilde g=0$.
  
  In the case of self-interaction also, if the Bose star is formed with net angular momentum, it could be unstable and the instability time scale can be quite smaller in comparison with   $\tilde{\tau}_{gr}$. 
 Figure \eqref{plot5} shows three snapshots of $|\psi|^2$ together with the vorticity magnitude at time $\tilde{t}\sim 16000$ in $xy$-plane for three different values of  $\mathcal{\tilde{L}}_{\rm tot}$ . The vorticity amplitude is shown by the black contours and whereas the bright red spots show the region where the Bose star has formed. The plots show that black contours are concentrated outside the star-forming region.
 Thus, in this case also like $\tilde{g}=0$ case, we believe the formed star does not have any intrinsic angular momentum ,and our results show stable stars. The results presented here are consistent with the stability analysis given in \cite{Levkov_2018}.

	\subsubsection{Repulsive self-interaction}

 Next, we consider the case of repulsive self-interaction. In this case, it is reasonable to expect that the bound state formed in the numerical results is due to gravity. If one increases the value of the coupling constant to values higher than 4.56, the bound states still can be formed.  
 But the study of such bound states requires higher computational time than allowed by the convergence test implemented for checking the validity of the numerical results. Keeping this restriction in mind, at present we restrict ourselves to $\tilde{g}=+4.56$.

 We first analyze the plots of maximum density $\max |\tilde{\psi}|^2$ vs. time as shown in Figure (7a). The topmost case corresponds to the case when the total angular momentum of the collapsing cloud is zero. The middle and the plot at the bottom respectively correspond to the cases with $\mathcal{\tilde{L}}_{\rm tot}\,=\,3\ \&\,5$.  All the plots shows that  $\max |\tilde{\psi}|^2$ remains small and constant till time $\tilde{t}\sim\tilde{\tau}_{gr}$.
 After that $\max |\tilde{\psi}|^2$ maximum density rises with time, till the time convergence test allows the program to run.
 Table \eqref{tabV} shows the estimated values of $\tilde{\tau}_{gr}$ for different values of $\mathcal{\tilde{L}}_{\rm tot}$. The table shows that value of $\tilde{\tau}_{gr}$  increases with increasing $\mathcal{\tilde{L}}_{\rm tot}$. 
  If one compares these values of $\tilde{\tau}_{gr}$ reported in Table\eqref{tabV} for different $\mathcal{\tilde{L}}_{\rm tot}$ with the corresponding value of  $\tilde{\tau}_{gr}$ shown in Table \eqref{tab1}, one can notice that the presence of repulsive interaction increases the gravitational condensation time. This is expected since the repulsive interaction between the constituent particles of the collapsing cloud can work with the quantum pressure to counter the effect of self-gravity. 
 
	

Next, Figure (7b)  depicts the plots of spectrum $\tilde{F}(\tilde{t},\tilde{\omega})$ versus frequency $\tilde{\omega}$ at time $\tilde{t}\simeq\, 25500$ for different values of $\mathcal{\tilde{L}}_{\rm tot}$. The presence of a peak at negative frequency in $\tilde{F}(\tilde{t},\tilde{\omega})$ implies the presence of a gravitationally bound state. 
The stability analysis in \cite{{Dmitriev_2021}} shows that an isolated Bose star with angular momentum might be stable. Therefore, it is of interest to know what happens to the net angular momentum of the collapsing cloud after the star formation. Figure (7a) shows three snapshots of density profiles together with the vorticity amplitude when $\mathcal{\tilde{L}_{\rm tot}}=5$. Condensation time when $\mathcal{\tilde{L}_{\rm tot}}=5$ is around 18000. The first snapshot on the left is for time $\tilde{t}\sim\,22000$ which is closer to  $\tilde{\tau}_{gr} \sim 18000$. In this plot, the black contours represent the presence of finite vorticity (angular momentum) outside the star forming region represented by the red spot. The middle snapshot is taken at time  23300 which shows the reduction in the black contours around the star forming region. This reduction is not just limited to $xy$-plane, in the other two planes also a strong  reduction in vorticity is seen out side the star-forming region. The third snapshot on the right is for time 27400 shows a near absence of vorticity outside the star-forming region. Here also vorticity is not seen in the other planes outside the star-forming region. Similar behavior is seen for  $\mathcal{\tilde{L}_{\rm tot}}=3$ case also that soon after the gravitational condensation the vorticity outside the star-forming region starts decreasing.  Table\eqref{tab VI} shows values of central density, the radius of the star $R_{95}$ and the mass of the star at time around 26500 for different values of $\mathcal{\tilde{L}_{\rm tot}}$. Now consider the cases when the collapsing cloud has nonzero angular momentum.  According to Table \eqref{tabV} the star formation for $\mathcal{\tilde{L}_{\rm tot}}=5$ happens significantly later than $\mathcal{\tilde{L}_{\rm tot}}=3$  case. Table \eqref{tabV} shows that this late-born star is not only approximately as massive as its predecessor but it is more compact also.  As we have seen before for $\tilde g=0$ and $ - 4.56$, in the present case also, as the central density of a Bose star increases it becomes  more compact.

\section{Conclusions}\label{sec4}

We have studied Bose star formation in the kinetic regime when the initial self-gravitating cloud has net angular momentum. In the earlier studies of Bose star formation, the role of angular momentum in the initial cloud was not considered \cite{Levkov_2018, chen:2021}.
For the case of an attractive self-interaction, we have shown that the coupling constant $\tilde{g}$ cannot be increased arbitrarily to form gravitationally bound structures like a Bose star. In this case, the coupling constant is required to satisfy the condition \eqref{bound}.  We have also shown that the property of boson star for $\tilde{g}\,\leq\,0$ and $\tilde{g}\,>0$ can be different. For the cases when $\tilde{g}\,\leq\,0$ the rotating 
 boson stars are known to be unstable  \cite{Dmitriev_2021}. The time scale associated with the instability is very small in comparison with the gravitational condensation time. The persistence of minimum in the power spectrum  $\tilde{F}(\tilde{t},\tilde{\omega})$ for time $\tilde{t}\,>\tilde{\tau}_{gr}$ together with analysis of vorticity amplitude shows the presence of  a stable and non-rotating Bose-star in our numerical results. In the case of repulsive interaction, we show that vorticity outside the star-forming region is absent and therefore the formed star might be having an intrinsic angular-momentum.
We  also show for a gravitationally bound structure to form in case of attractive self-interaction  there exists a bound described by condition \eqref{bound}. It is also shown that the introduction of net angular-momentum in the initial self-gravitating cloud can not only significantly  alter the gravitational condensation time, but also affects the compactness(radius $R_{95}$) and mass of the star.

\section{Acknowledgments}\label{sec5}
We are grateful to  Bodo Schwabe, Mateja Gosenca, and Dilip Angom for helpful discussions and helping with the publicly available code  \texttt{AxioNyx}. All the computations are accomplished on the Vikram-100 HPC cluster at Physical Research Laboratory, Ahmedabad.

\end{document}